\documentclass[conference]{IEEEtran}
\IEEEoverridecommandlockouts
\usepackage{cite}
\usepackage{amsmath,amssymb,amsfonts}
\usepackage{algorithmic}
\usepackage{graphicx}
\usepackage{textcomp}
\usepackage{xcolor}
\def\BibTeX{{\rm B\kern-.05em{\sc i\kern-.025em b}\kern-.08em
    T\kern-.1667em\lower.7ex\hbox{E}\kern-.125emX}}

\usepackage{fancyhdr}
\begin{document}

\title{Towards Ontology-Based Requirements Engineering for IoT-Supported Well-Being, Aging and Health\\
}

\author{\IEEEauthorblockN{Hrvoje Belani}
\IEEEauthorblockA{\textit{Directorate for e-Health} \\
\textit{Ministry of Health}\\
Zagreb, Croatia \\
hrvoje.belani@miz.hr}
\and
\IEEEauthorblockN{Petar Šolić}
\IEEEauthorblockA{\textit{Dept. of Electronics and Computing} \\
\textit{University of Split, FESB}\\
Split, Croatia \\
psolic@fesb.hr}
\and
\IEEEauthorblockN{Toni Perković}
\IEEEauthorblockA{\textit{Dept. of Electronics and Computing} \\
\textit{University of Split, FESB}\\
Split, Croatia \\
toperkov@fesb.hr}
}

\maketitle
\thispagestyle{fancy}
\begin{abstract}
Ontologies serve as a one of the formal means to represent and model knowledge in computer science, electrical engineering, system engineering and other related disciplines. Ontologies within requirements engineering may be used for formal representation of system requirements. In the Internet of Things, ontologies may be used to represent sensor knowledge and describe acquired data semantics. Designing an ontology comprehensive enough with an appropriate level of knowledge expressiveness, serving multiple purposes, from system requirements specifications to modeling knowledge based on data from IoT sensors, is one of the great challenges. This paper proposes an approach towards ontology-based requirements engineering for well-being, aging and health supported by the Internet of Things. Such an ontology design does not aim at creating a new ontology, but extending the appropriate one already existing, SAREF4EHAW, in order align with the well-being, aging and health concepts and structure the knowledge within the domain. Other contributions include a conceptual formulation for Well-Being, Aging and Health and a related taxonomy, as well as a concept of One Well-Being, Aging and Health. New attributes and relations have been proposed for the new ontology extension, along with the updated list of use cases and particular ontological requirements not covered by the original ontology. Future work envisions full specification of the new ontology extension, as well as structuring system requirements and sensor measurement parameters to follow description logic.
\end{abstract}


\section{Introduction}
Throughout 2020, 2021 and 2022 the global COVID-19 pandemic has impacted all aspects of people's well-being, with health being one of them. As reported by the Organisation for Economic Cooperation and Development (OECD) \cite{b1}, the new SARS-CoV-2 coronavirus has had devastating impacts on both physical health and mortality, with averaged 16\% excess deaths in 33 OECD countries between March 2020 and early May 2021 compared to 2015-2019 period, which resulted in a 7-month fall in OECD 29-average life expectancy in 2020. Also, mental health deteriorated for almost all population groups on average in 2020, with older people much more likely suffering severe outcomes or death due to the infection, making reduced social contact an especially important precaution for them, while younger adults have experienced some of the largest declines in mental health, social connectedness and life satisfaction both in 2020 and 2021.

Confronting such a global, not only public health but also economic and social crisis, with both short and potentially long term consequences, showed all the seriousness of dealing with well-being, aging, and health (WBAH) as important aspects of life that affect everybody \cite{b2}. To inspect definitions of each of the WBAH aspects, health is defined by the World Health Organization (WHO) as "a state of complete physical, mental and social well-being and not merely the absence of disease and infirmity" \cite{b3}. As a process of getting older, aging represents the accumulation of changes in a human being over time and can encompass physical, psychological, and social changes \cite{b4}. The OECD defines current human well-being as a term encompassing the different areas that matter for people’s lives and covering well-being outcomes at the individual, household or community level, focusing on material conditions, quality of life factors and community relations through the following 11 dimensions of The OECD Well-Being Framework: Income and Wealth, Work and Job Quality, Housing, Health, Knowledge and Skills, Environment Quality, Subjective Well-Being, Safety, Work-Life Balance, Social Connections and Civil Engagement \cite{b5}.

WBAH aspects are being addressed at the policy level by recent global initiatives. The United Nations (UN) Decade of Healthy Ageing (2021-2030) is a global collaboration, aligned with the last ten years of the UN Sustainable Development Goals (SDGs), aiming to foster healthy aging and improve the lives of older people and their families and communities, by addressing four areas for action: (1) Age-friendly environments; (2) Combating ageism; (3) Integrated care, and (4) Long-term care \cite{b6}. In order to respond to mental health challenges risen during COVID-19 pandemic, the WHO Comprehensive Mental Health Action Plan 2013–2030 \cite{b7} has been updated with implementation options and indicators, and the WHO European Framework for Action on Mental Health 2021-2025 has been endorsed, providing "a coherent basis for intensified efforts to mainstream, promote and safeguard mental well-being as an integral element of COVID-19 response and recovery; to counter the stigma and discrimination associated with mental health conditions; and to advocate for and promote investment in accessible quality mental health services" \cite{b8}. WBAH is a complex concept of interleaving terms, underlying aspects and containing dimensions.

This paper proposes an approach towards ontology-based requirements engineering for well-being, aging and health supported by the Internet of Things (IoT). After a research literature review on usage of ontologies for both requirements engineering discipline and implementation of IoT solutions, the appropriate ontology standard has been identified for further extension in order to cover the WBAH aspects and dimensions.

The paper is organized as follows: Section II. presents the motivation and related work on knowledge-based RE, ontologies for RE and ontologies for IoT. Section III. provides a WBAH taxonomy outline for IoT and a mapping of established areas of Ambient Assisted Living, Active and Health Aging, and Health and Well Being with WBAH concepts and dimensions. It also proposes a conceptual formulation for WBAH. Section IV. presents IoT-based RE-ready WBAH ontology design. Section V. provides discussion on the usefulness and limitations of the proposed approach, and its possible impact outside the WBAH scope. Also, a concept of "One Well-Being, Aging and Health" has been introduced. Section VI. provides a conclusion with directions for future work in building IoT-based ontologies for WBAH, that can inform requirements specifications for future innovative IoT services.

\section{Motivation and Related Work}
In order to better cope with the negative effects of COVID-19 pandemic and address WBAH aspects properly, new systems, products and services are needed, for which requirements have to be crafted in a multidisciplinary and cross-domains manner. Requirements engineering for WBAH (REWBAH) can e.g. "identify specific challenges of uncertainty, anxiety, and loneliness from different age perspectives, and help assess how emerging home-office/mixed-office environments influence the well-being and health of people" \cite{b2}.

Requirements engineering (RE) is a discipline traditionally belonging to firstly software engineering and then systems engineering. RE seeks for a domain knowledge in order to provide factual and rich vocabulary for expressing needs and wants of the new system and service users, as well as interacting with other systems and the everchanging environments. Therefore, one of the recognized challenges of RE is knowledge integration. Furthermore, domain knowledge needs to be automatically processed, mapped to other known schemes and linked to data in order for RE to gain full benefit of it. So, the second important challenge of RE is automatic processing support.

In order to address the above mentioned challenges, RE as the other software engineering disciplines need the support of knowledge engineering (KE) methods. This field is certainly not new and has been widely researched for more than a three decades \cite{b9}, addressing various aspects \cite{b10} and critical factors \cite{b11} of knowledge-based software engineering \cite{b12} \cite{b13}. When focusing on the earliest phases of software development, dealing with uncertain and inconsistent requirements becomes even greater challenge to be tackled by some of the KE methods.

\subsection{Knowledge-Based Requirements Engineering}
One of the earliest works \cite{b14} in the efforts to provide knowledge-based support for RE has targeted requirements specification as the process that "should be considered from a viewpoint that is close to an analyst's cognitive processes" and proposes the support environment that "exploits knowledge-based paradigms for the capture and modelling of facts about an application domain, which are then transformed into a functional specification". In order to develop a proper requirements specification, there is one more type of requirements analysis knowledge needed besides domain knowledge; the method knowledge, related to particular development approach depending on the system under development. It is claimed that both types of knowledge "need to be captured and accessed by a requirements specification tool so that more accurate and flexible specifications may be developed" \cite{b15}. This paper aims at establishing an ontology for WBAH which may also serve as a knowledge base for further requirements specification for the particular WBAH system, product or service planning to be developed.

Analysis and modeling for a specific domain target at a family of systems rather than a single system under development. They aim for system requirements and architectures reuse in order to achieve more efficient and flexible system development. Such a knowledge-based approach that enables instantiating a target system specification from the domain model asks for an application-domain independent prototype environment that provides "rules for generating target system specifications from the domain model" \cite{b16}. In order to reuse not only system requirements already formulated but the underlying knowledge acquired during requirement elicitation, modeling and validation activities, the common representation of domain knowledge is needed where "requirement knowledge allows exploitation by common tools" and "the developer is able to browse a populated domain knowledge model alongside the building and population of a new requirement model" \cite{b17}. These findings also support the efforts towards creating WBAH ontology that will enable requirements engineers reuse already acquired domain knowledge, which seems of great importance given the complexity of WBAH environments.

In the following few decades significant research efforts have been made to address different knowledge-based engineering aspects to improve traditional RE. The role of domain knowledge representation in requirements elicitation has been addressed stating "not only the contents but also the representation is important" and defining and comparing various metrics to evaluate the effect of the representations on the quality of requirements specifications \cite{b17}. The framework for knowledge-based RE supporting automatic detection of a range of inconsistencies between requirements has been proposed \cite{b18}. It has used description logic to capture requirements and form fundamental logical system for requirements analysis and reasoning, even applying it to ontologies in order to provide precise meanings to the terms used to specify requirements \cite{b19}.

\subsection{Ontologies for Requirements Engineering}
As originally defined back in 1992 for the areas of computer and information sciences \cite{b20}, ontology defines a set  of "classes (or sets), attributes (or properties), and relationships (or relations among class members) with which to model a domain of knowledge or discourse" and includes "information about their meaning and constraints on their logically consistent application" \cite{b21}. In the context of the RE process, an ontology can be created and constantly updated to "store the definitions and relationships of those classes, properties, and individuals to enable the analysis and reasoning on the requirements" \cite{b21}.

In order for requirements engineers and analysts to gain an understanding to the system's domain, environment, and the system’s stakeholders \cite{b22}, domain understanding and requirements elicitation have to take place. To provide an efficient reuse of ontologies, they are usually divided to: (1) generic (top-level, foundation, core) ontologies - contain knowledge and domain-independent concepts that can be reusable across various domains; (2) specific ontologies - represent concepts in a way that is specific to a particular domain, application, task, activity, method, etc. Specific ontologies are further divided to: (2.1) domain ontologies - represent concepts of a particular domain (e.g. medicine); (2.2) task ontologies - contain the terms and a task or an activity (e.g. diagnosing a patient); (2.3) application ontologies - contain subsets of concepts from domain and task ontologies intended to be used in a specific application \cite{b23}.

Some authors \cite{b24} claim that "application ontology construction must be of responsibility of the requirements engineering team" and "requirements engineers should be prepared to produce such application ontologies, as software designers will have to learn on how to best use these application ontologies". Classification of approaches that include ontologies within RE \cite{b25} consists of the ontologies for: (1) describing requirements specification documents - to capture the RE documents structures and adapt the same content in diverse formats in order to be understandable by all stakeholders and reutilized in diverse projects; (2) formally representing requirements - to present a semantic structure for capturing requirements relevant information, in order to support the RE process semantically (e.g. KAOS methodology); (3) formally representing application domain knowledge - to provide systematization for RE elicitation, model and analysis of ontology terms by using the Language Extended Lexicon (LEL) in order for the system's use context to be understood in detail before requirements can be derived.

The systematic review identified the primary studies on the use of ontologies in RE \cite{b26}, finding there are empirical evidences to state that ontologies benefit RE activities in both academy and industry settings, helping to reduce ambiguity, inconsistency and incompleteness of requirements and providing tool support and usage of W3C recommended ontology languages. Various ontology-based frameworks \cite{b27}, models \cite{b28} and methods \cite{b29} have been proposed in order to support different RE phases, some even building a domain ontology for requirements classification, so-called Requirements Classification Ontology (RCO) \cite{b30}. Whichever purposes and RE phases the proposed ontology serves, it should be designed by using a proper ontology engineering method (e.g. Methontology) and provided with a tool support (e.g. Protégé), in order to verify and validate the developed ontology.

\subsection{Ontologies for Internet of Things}
Besides representing concepts and describing knowledge in RE, ontologies can be used within specific technological areas, like Internet of Things (IoT), in order to represent sensor knowledge and describe acquired data semantics. Multimodality of sensors and the usage of cross-domain IoT platforms shape heterogeneity which poses interoperability and design challenges and limits the possibility of reusing sensor data to develop new applications and integrating automated solutions based on sensor data. In order to tackle these challenges, the study from 2017 shown "no existing ontology is comprehensive enough to document all the concepts required for semantically annotating an end-to-end IoT application as ontologies are often restricted to a certain domain" and "there are no concrete methods for evaluating an ontology, and developers must always follow the best practices while publishing a new ontology in order to enhance readability, usability, extensibility, and interoperability" \cite{b31}.

With a growing adoption of IoT in different fields, such as environment monitoring, healthcare services, transportation and logistics and smart homes and cities, a variety of numerous ontologies has been developed and eventually collected in the Linked Open Vocabularies for Internet of Things (LOV4IoT) ontology catalog for IoT \cite{b32}, with currently 802 ontology-based IoT projects listed in and 29 IoT application areas analyzed. 289 of these ontologies (36\%) align with the WBAH aspects and belong to the following 10 LOV4IoT application areas: Home (63), Emotion (49), Food (53), Health (79), Depression (6), Ambient Assisted Living (11), Activity (11), Wearable (6), Fitness (5), Air Quality (6).

Some of the recent research on IoT ontologies delivered the unified ontology aiming to achieve semantic interoperability among heterogeneous testbeds \cite{b33}; the IoT-Stream lightweight ontology for data stream analytics and event detection \cite{b34}; the COIoT comprehensive ontology by reusing core concepts from existing ontologies and complementing concepts like policy, context, services and monitoring \cite{b35}; the evaluation method for ontologies verification and validation by using evaluation tools that detect errors by diagnosing various metrics and pitfalls \cite{b36}; the IoT architecture for virtual IoT devices and their semantic framework deployed at the edge of network and able to aggregate capabilities of IoT devices and derive new services by inference \cite{b37}.

Finally, the IoT ontology landscape report from December 2021 \cite{b38} included 30 ontologies from different application areas of IoT, assessing "the technology readiness level (TRL), ranging from technology validated in the lab (very light) to actual system proven in operational environment (dark)" for each ontology. Six ontologies are listed as generic (horizontal), while others fitted in one or more of the following application areas: Home/Building (6), Industry (6), Mobility (3), Health (1), Energy (3), Cities (3), Wearables (2), Farming/Agrifood (4) and Water/Environment (3). Not many IoT ontologies assessed align with the WBAH aspects, only these in Health, Wearables and partly Water/Environment application areas.

\subsection{IT vs. OT in Internet of (Medical) Things}
For efficient implementation of IoT into WBAH solutions, it is important to distinguish among the complexity options of envisioned solution in terms will it implement Industrial Internet of Things (IIoT) or Industrial Internet of Medical Things (IIoMT) \cite{b39} \cite{b40} even more. IIoT is defined as "a system comprising networked smart objects, cyber-physical assets, associated generic information technologies (IT) and optional cloud or edge computing platforms, which enable real-time, intelligent, and autonomous access, collection, analysis, communications, and exchange of process, product and/or service information, within the industrial environment, so as to optimise overall production value. This value may include: improving product or service delivery, boosting productivity, reducing labour costs, reducing energy consumption, and reducing the build-to-order cycle." \cite{b41}.

IIoT differs from IoT in the following prospects \cite{b42}: (1) Linked things - from affordable user-level devices to costly machines, sensors, systems, with high degree of difficulty; (2) Service model - from human-based to machine-based; (3) Communication capacity - from a smaller number of communication standards to a large range of connectivity technologies and standards; (4) Communication transportation - from typically wireless to both wired and wireless; (5) Amount of data - from medium-high to high-very high; (6) Evaluative - from quite trivial to serious, in terms of timing, security, privacy, reliability, etc. IIoT systems align with the well-established concept of Industrial Automation and Control Systems (IACS), often referred to as Operational Technology (OT). IIoT represents a showcase of converging, aligning and integrating of IT and OT environments.

Application of IIoT to medical science and practices, as well as healthcare, even big pharma, biomanufacturing and life sciences in general, is named IIoMT and includes IoT medical devices in IIoT settings for biomedical, medical and health related purposes. With the certain degree of automation, intelligence and autonomy, when applied to healthcare and medicine it can achieve a full potential of the terms "smart healthcare" and "smart medical care". In the same time it is posing new challenges by widening the critical requirements space, especially for non-functional ones, such as high performance, cybersecurity, operational and patient safety, medical device reliability, etc. \cite{b43}.

\section{Developing WBAH Taxonomy for IoT}
In order to unlock the full potential for development of innovative IoT solutions for WBAH, an ontology is needed, comprehensive enough to cover the WBAH aspects, represent the captured knowledge and enable full interoperability of sensor networks, not only semantic but also process-wise. Furthermore, the same ontology may serve to requirements engineers for shaping specification documents for new IoT-supported WBAH services. Firstly, a conceptual formulation of WBAH is needed, to serve as an input for developing WBAH taxonomy for IoT. Then, this taxonomic structure will serve as a backbone \cite{b23} of the proposed IoT-based RE-ready ontology for WBAH. The new ontology development process may include selecting one or more ontologies already developed to describe the same topic or a part of domain, but this may not satisfy all the requirements that have to be fulfilled, and one of the great challenges is knowledge maintenance, since all parts of knowledge, including ontological knowledge, evolve over time \cite{b44}.

\subsection{Definitions}
\begin{figure}[htbp]
\centerline{\includegraphics[width=60mm]{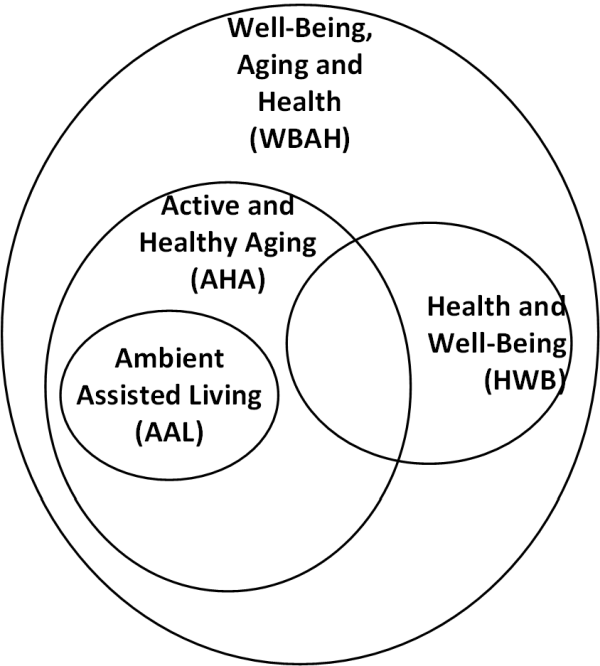}}
\caption{Venn diagram for AAL, AHA, HWB and WBAH.}
\label{aal_aha_hwb_wbah}
\end{figure}

Before delivering a conceptual formulation of WBAH, the interrelationships with already established areas need to be addressed, namely: Ambient Assisted Living (AAL), Active and Healthy Aging (AHA) and Health and Well-Being (HWB). Fig.~\ref{aal_aha_hwb_wbah} illustrates these interrelationships by using Venn diagram.

AAL or innovative ICT-enabled assisted living relates to intelligent systems of assistance for a better, healthier and safer life in the preferred living environment and covers concepts, products and services that interlink and improve new technologies and the social environment, with the aim of enhancing the quality of life (related to physical, mental and social well-being) to for all people (with a focus on older persons) in all stages of their life \cite{b45}.

AHA is a complex and multi-dimensional concept with marked heterogeneity, with the conceptual AHA framework that includes three key domains (1. Physical and cognitive capability across the life course; 2. Psychological and social well-being, mental health and quality of life across the life course; 3. Functioning of underlying physiological systems across the life course, preventing or delaying onset of chronic diseases, frailty and disability) and three key influencing factors (4. Education, lifelong learning, working and caring; 5. Lifetime lifestyles; 6. Lifetime social, economic and physical environment) \cite{b46}.

"HWB comprise physical health, psychological and emotional stability, and social engagement. Physical wellness involves self-care and a temperate lifestyle. Emotional well-being is psychological well-being encompassing subjective experience and positive emotionality. A stable mood—emotional equanimity—enhances countering negative emotions and physician burnout. Social engagement revolves around interpersonal and social relations. Physician engagement entails a doctor's commitment to studying, enhancing expertise, and skills toward safe and high-quality patient care." \cite{b47}.

Based on the definitions provided and the previous research referenced, the list of WBAH dimensions can be assembled and each of the dimensions mapped against the areas of AAL, AHA and HWB. The Table~\ref{table_WBAH} shows WBAH dimensions coverage by AAL, AHA and HWB, providing three recognized levels of coverage: full coverage (FC), partial coverage (PC) or no coverage (NC). Each of the dimensions is attributed a type, depending if it belongs to well-being, aging or health aspect.

 For the purpose of further ontology development, the following conceptual formulation for WBAH is proposed: "WBAH is a multidimensional concept covering well-being, aging and health aspects of persons, communities and societies in the appropriate contexts including the environment properties, the entities engagement and interactions, as well as past and current WBAH status of each of the entities observed in order to achieve and maintain WBAH at the satisfied level and gain resilience against factors that pose risk to degrade WBAH".

\begin{table}[htbp]
\caption{WBAH dimensions coverage by AAL, AHA and HWB}
\begin{center}
\begin{tabular}{|c|c|p{3.0cm}|c|c|c|}
\hline
\textbf{No.} & \textbf{\textit{Ty.}}& \textbf{\textit{WBAH dimensions}}& \textbf{\textit{AAL}}& \textbf{\textit{AHA}}& \textbf{\textit{HWB}} \\
\hline
1. & WB & Income and wealth & NC & NC & NC \\
\hline
2. & WB & Work and job & PC & FC & FC \\
\hline
3. & WB & Quality & PC & FC & FC \\
\hline
4. & WB & Housing & FC & FC & PC \\
\hline
5. & WB & Health & PC & FC & FC \\
\hline
6. & WB & Knowledge and skills & NC & PC & PC \\
\hline
7. & WB & Environment quality & FC & FC & FC \\
\hline
8. & WB & Subjective well-being & NC & PC & FC \\
\hline
9. & WB & Safety & FC & FC & NC \\
\hline
10. & WB & Work-life balance & FC & FC & FC \\
\hline
11. & WB & Social connections and civil engagement & FC & FC & PC \\
\hline
12. & A & Functional aging: social & FC & FC & PC \\
\hline
13. & A & Functional aging: psychological & PC & FC & FC \\
\hline
14. & A & Functional aging: physiological & FC & FC & FC \\
\hline
15. & A & Biological aging & PC & PC & PC \\
\hline
16. & A & Chronological aging & PC & PC & FC \\
\hline
17. & H & Physical health & FC & FC & FC \\
\hline
18. & H & Mental health & PC & FC & FC \\
\hline
19. & H & Health outcomes: personal health outcome & FC & FC & FC \\
\hline
20. & H & Health outcomes: healthcare outcome & FC & FC & FC \\
\hline
21. & H & Health outcomes: public health outcome & PC & FC & FC \\
\hline
22. & H & Determinants of health: personal & FC & FC & FC \\
\hline
23. & H & Determinants of health: social & PC & FC & FC \\
\hline
24. & H & Determinants of health: ecological & FC & FC & PC \\
\hline
\multicolumn{6}{l}{$^{\mathrm{a}}$Legend: FC - fully covered;  PC - partially covered; NC - not covered.}
\end{tabular}
\label{table_WBAH}
\end{center}
\end{table}

As it can be seen, AAL is focused on usage of technologies in the environments, AHA is targeting aging and HWB is more oriented towards well-being of individuals in personal and professional surroundings, while WBAH is aiming to comprehensively include all aspects of well-being, aging and health.

\subsection{Research Methods}

In order to analyse the perspective of establishing an ontology with dual purpose - for informing requirements specification process and for describing knowledge structured through the IoT data gathering, the secondary data analysis has been conducted and presented in Section II. Given the fuzzy boundaries around the WBAH concept, the conceptualization of WBAH has been formulated in Section III. and the mapping of AAL, AHA and HWB areas to each of the recognized WBAH dimensions proposed. This mapping may serve as a starting point for more comprehensive scoping review of AAL, AHA and HWB areas in order to synthesize research evidence, categorize existing literature and align it with each of the WBAH dimensions.

The following Section IV. provides a short analysis of existing core IoT ontologies in order to assess their alignment to WBAH concepts and dimensions. The aim is to extend the ontology which is better fit to serve as a baseline ontology for alignment with the WBAH concepts. However, this process should not exclude other analysed ontologies because there is always a possibility that some concepts will be needed in further development of a new WBAH ontology, which is out of the scope of this paper.

The rest of the Section IV. provides an analysis of use cases and requirements specified by a global standardisation body in order to inform the design of an IoT ontology within the domain compatible with WBAH, but still missing some WBAH concepts and dimensions. This work contributes the efforts with the updated list of use cases not covered by the original ontology, derived from existing ones as well as extracting new use cases from the recognized WBAH dimensions.

\section{IoT-based RE-ready WBAH Ontology Design}
Ontology design have to firstly be informed about the ontology purpose. The aim of this paper is to design a domain ontology that provides semantic interoperability for the IoT environment and also enables building of requirements specifications for WBAH solutions that use sensor networks in supporting entities and their interactions in the shared environment. Given the related work, starting point for IoT-based RE-ready WBAH ontology design should not not aim at creating a new ontology, but extending the appropriate one already existing, in order to align with the WBAH concepts and structure the knowledge within the domain.

\subsection{W3C/OGC SSN and ETSI SAREF Ontologies}
Back in 2012 the World Wide Web Consortium (W3C) started working on the Semantic Sensor Network (SSN) ontology and published it as a W3C Recommendation, as well as an OGC implementation standard, in cooperation with the Open Geospatial Consortium (OGC). SSN is "an ontology for describing sensors and their observations, the involved procedures, the studied features of interest, the samples used to do so, and the observed properties, as well as actuators" \cite{b48}. Based on a modular architecture, SSN also includes a self-contained core ontology named SOSA (Sensor, Observation, Sample, and Actuator) for its elementary classes and properties. 

SSN and SOSA set of ontologies "support a wide range of applications and use cases, including satellite imagery, large-scale scientific monitoring, industrial and household infrastructures, social sensing, citizen science, observation-driven ontology engineering, and the Web of Things" \cite{b49}. The latest corrections to the SSN ontology were made in 2017. Some shortcomings of SSN include real-time data collection issues, providing a taxonomy for measurement units, context, quantity kinds (the phenomena sensed), and exposing sensors to services \cite{b31}.

Another major effort in IoT ontology standardisation is made by the European Telecommunications Standards Institute (ETSI) in 2020, by introducing the Smart Applications REFerence (SAREF) ontology, "intended to enable interoperability between solutions from different providers and among various activity sectors in the Internet of Things (IoT), thus contributing to the development of the global digital market" \cite{b48}. It is a core IoT ontology currently offering extensions for 10 domains (Energy, Environment, Building, Smart Cities, Industry and Manufacturing, Smart Agriculture and Food Chain, eHealth/Ageing-well, Wearables, Water, Smart Lifts), while the extension for the Automotive domain is under development.

\subsection{SAREF4EHAW Ontology Extension}

Two of SAREF ontology extensions partially aligned with WBAH concepts are SAREF4health from 2018 and more comprehensive SAREF4EHAW published in 2020. The objective of SAREF4EHAW is to extend SAREF ontology for the eHealth/Ageing-well (EHAW) vertical by "investigating EHAW domain related resources, as reported in ETSI TR 103 509 \cite{b50}, such as: potential stakeholders, standardization initiatives, alliances/associations, European projects, EC directives, existing ontologies, and data repositories" \cite{b51}. SAREF4EHAW is an OWL2-DL ontology, using the Web Ontology Language v2 (OWL2) for authoring ontologies, which belongs to Description logics (DL) family of formal knowledge representation languages.

In order to further justify the SAREF4EHAW ontology extension towards a new WBAH ontology, it is important to point out that SAREF reference ontology "proposes basic functions that can be combined in order to have more complex functions in a single device" and "a Device offers a Service which is a representation of a Function to a network that makes the function usable by other devices in the network" \cite{b51}. SAREF has also been mapped with oneM2M base ontology in 2017, which "describes key classes, relations and properties that are necessary to enable semantic functionalities and interoperability between applications" \cite{b50}.

\begin{figure}[htbp]
\centerline{\includegraphics[width=95mm]{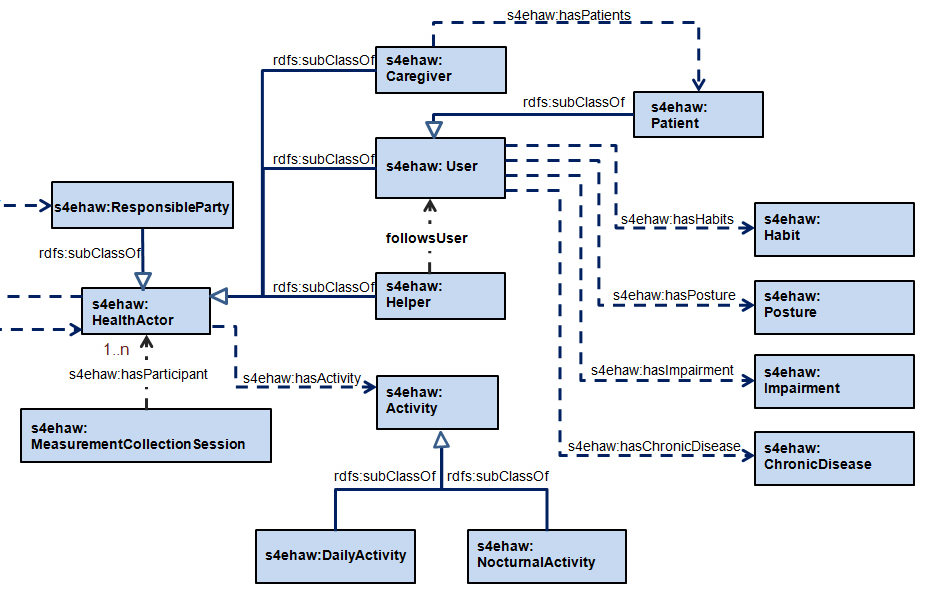}}
\caption{Part of the semantic model for SAREF4EHAW ontology \cite{b58}.}
\label{saref4ehaw}
\end{figure}

SAREF4EHAW can be mainly described by the following self-contained knowledge sub-ontologies or modules: HealthActor, Ban, HealthDevice, Function (measured data related concepts included) and Service. The part of a high level view of the semantic model for HealthActor is shown in Fig.~\ref{saref4ehaw}. Therefore, this ontology is a solid candidate to be further extended to support WBAH concepts, by introducing new additions particular to each of the WBAH dimensions, as well as listing new requirements to be specified and supported by the proposed extension. Some existing SAREF extensions may also contribute to the particular WBAH dimension with their sub-ontologies.

\begin{table}[htbp]
\caption{WBAH dimensions coverage by SAREF and extensions}
\begin{center}
\begin{tabular}{|c|c|p{2.25cm}|p{4cm}|}
\hline
\textbf{No.} & \textbf{\textit{Ty.}}& \textbf{\textit{WBAH dimensions}}& \textbf{\textit{SAREF extension or proposed}} \\
\hline
1. & WB & Income and wealth & User is in domain of s4wbah:feelsPositive and has members s4wbah:confident, s4wbah:good, s4wbah:inspired, s4wbah:interested, s4wbah:proud
User is in domain of s4wbah:feelsNegative and has members s4wbah:afraid, s4wbah:angry, s4wbah:bad, s4wbah:bored, s4wbah:sad, s4wbah:stressed, s4wbah:upset \\
\hline
2. & WB & Work and job & SAREF4INMA \\
\hline
3. & WB & Quality & SAREF4AGRI \\
\hline
4. & WB & Housing & SAREF4BLDG, SAREF4LIFT \\
\hline
5. & WB & Health & SAREF4EHAW \\
\hline
6. & WB & Knowledge and skills & SAREF4INMA  \\
\hline
7. & WB & Environment quality & SAREF4ENVI \\
\hline
8. & WB & Subjective well-being & User is in domain of s4wbah:isHappy and has members
s4wbah:notaveryhappy, s4wbah:lesshappythanpeers, s4wbah:generallynotveryhappy, s4wbah:somewhathappy, s4wbah:generallyhappy, s4wbah:morehappythanpeers, s4wbah:veryhappy \\
\hline
9. & WB & Safety & SAREF4ENER, SAREF4AUTO \\
\hline
10. & WB & Work-life balance & Same as 1. \\
\hline
11. & WB & Social connections and civil engagement & SAREF4EHAW \\
\hline
12. & A & Functional aging: social & User is in the domain of s4wbah:hasCompanion and has members s4wbah:dog, s4wbah:cat, s4wbah:fish, s4wbah:reptile, s4wbah:hamster, s4wbah:rabbit, s4wbah:parrot, s4wbah:otherbird, s4wbah:otherpet \\
\hline
13. & A & Functional aging: psychological & SAREF4EHAW, same as 12. \\
\hline
14. & A & Functional aging: physiological & SAREF4WEAR \\
\hline
15. & A & Biological aging & SAREF4EHAW \\
\hline
16. & A & Chronological aging & SAREF4EHAW \\
\hline
17. & H & Physical health & SAREF4EHAW \\
\hline
18. & H & Mental health & SAREF4EHAW, same as 8. \\
\hline
19. & H & Health outcomes: personal health outcome & SAREF4WEAR \\
\hline
20. & H & Health outcomes: healthcare outcome & SAREF4EHAW \\
\hline
21. & H & Health outcomes: public health outcome & SAREF4CITY \\
\hline
22. & H & Determinants of health: personal & SAREF4EHAW \\
\hline
23. & H & Determinants of health: social & SAREF4EHAW, same as 8. \\
\hline
24. & H & Determinants of health: ecological & SAREF4WATR \\
\hline
\multicolumn{4}{l}{$^{\mathrm{a}}$SAREF extensions available at: https://saref.etsi.org/extensions.html.}
\end{tabular}
\label{table_SAREF4WBAH}
\end{center}
\end{table}

\subsection{Towards the WBAH Ontology Proposal}
In order to align WBAH dimensions listed in Table~\ref{table_WBAH} with some of the existing SAREF extensions, along with the SAREF4EHAW extension, WBAH taxonomy has been outlined containing dimensions derived from the well-being \cite{b52}, aging \cite{b53} and health \cite{b54} literature, ending up in proposing some new attributes and relations which begin composing the new WBAH ontology proposal, as shown in Table~\ref{table_SAREF4WBAH}. In order to better inform the ontology outline, the list of use cases (listed below as UC01-UC12) and ontological requirements specified for SAREF4EHAW \cite{b55} has been revisited. Firstly, the original list of SAREF4EHAW use cases (UC01-UC12) has been updated with the non-exhaustive list of new use cases (UC13-UC24), one for each of the WBAH dimensions:
\begin{itemize}
\item UC01: Elderly at home monitoring and support,
\item UC02: Monitoring and support of healthy lifestyles for citizens,
\item UC03: Early Warning System (EWS) and Cardiovascular Accidents detection,
\item UC04: Daily activity monitoring,
\item UC05: Integrated care for older adults under chronic conditions,
\item UC06: Monitoring assisted persons outside home and controlling risky situations,
\item UC07: Emergency trigger,
\item UC08: Exercise promotion for fall prevention and physical activeness,
\item UC09: Cognitive stimulation for mental decline prevention,
\item UC10: Prevention of social isolation,
\item UC11: Comfort and safety at home,
\item UC12: Support for transportation and mobility,
\item \underline{UC13:} Assessment of income and wealth satisfaction of an aging person,
\item \underline{UC14:} Monitoring of sanitary facilities access and usage,
\item \underline{UC15:} Assessment of happiness of an aging person,
\item \underline{UC16:} Risky behaviours monitoring, like smoking or alcohol intake,
\item \underline{UC17:} Activity monitoring for leisure and personal care, such as eating and sleeping,
\item \underline{UC18:} Participation promotion in accessible civic activities, like voting and consultations,
\item \underline{UC19:} Correct and timely drug intake assistance to avoid multi-drug interactions,
\item \underline{UC20:} Maintaining social contacts in the neighbourhood,
\item \underline{UC21:} Active participation in the household activities,
\item \underline{UC22:} Waste sorting assistance
\item \underline{UC23:} Monitoring interaction of an aging person with a pet companion,
\item \underline{UC24:} Assisting at education and skills activities in the community.
\end{itemize}

As it can be seen above, some of the UC13-UC24 use cases have been derived from the UC01-UC12 ones, e.g. U17 and UC21 from UC04; UC20 and UC23 from UC10. The reason for this is not having enough expressiveness in the existing SAREF4EHAW ontology to support the derived use cases, so the explicit use case specification is needed in order to ensure use case coverage by the new WBAH ontology being designed.

Total of 43 ontological requirements have already been recognized \cite{b51} for SAREF4EHAW extension, as well as 27 service requirements for IoT solutions (named "general service level assumptions" in the ETSI documentation) resulted from a reverse engineering process that has been previously carried out by the ETSI stakeholders, taking as input the initiatives and stakeholders from the UC01-UC12 use cases. Also, additional 59 service level assumptions have been listed in the ETSI documentation for specific sub-domains and associated use cases that might appear. These additional service level assumptions are grouped into the following categories \cite{b51}, which can direct further development of WBAH ontology:
\begin{itemize}
\item EHAW-DAM (Daily Activity Monitoring),
\item EHAW-UCC (Under Chronic Conditions),
\item EHAW-MOH (Monitoring Outside Home),
\item EHAWEMT (Emergency Trigger),
\item EHAW-EXP (Exercise Promotion),
\item EHAW-MDP (Mental Decline Prevention),
\item EHAW-PSI (Prevention of Social Isolation) and
\item EHAW-STM (Support for Transportation and Mobility).
\end{itemize}

Additional ontological requirements for WBAH proposal can now be defined from some of the new use cases (UC13-UC24), in order to ensure new ontology supporting the appropriate WBAH dimensions, e.g.:
\begin{itemize}
\item WBAH-01: Patient's (or assisted person's) happiness scale has to be modeled and its levels recognized in a timely manner. Combination of such an input with other sensor data may provide additional insight into the patient's (or assisted person's) next actions and possible risks.
\item WBAH-02: Patient's (or assisted person's) positivity scale has to be modeled and its levels recognized in a timely manner. Combination of such an input with other sensor data may provide additional insight into the patient's (or assisted person's) next actions and possible risks.
\item WBAH-03: Patient's (or assisted person's) negativity scale has to be modeled and its levels recognized in a timely manner. Combination of such an input with other sensor data may provide additional insight into the patient's (or assisted person's) next actions and possible risks.
\item WBAH-04: Patient, or assisted person, has regular interactions with his/her companion pet, and their interactions have to be modeled and monitored. Some of these activities, when combined with the happiness scale monitoring, confirm the well-being status of the patient. Also, when combined with the negativity scale it could point out to some risky behaviour.
\end{itemize}

\section{Discussion}
The presented approach for the WBAH ontology design builds upon the past high-quality work done by the ontology communities and standardisation bodies, enabling further WBAH extensions of the appropriate existed ontology (SAREF4EHAW), which already represents an extension of the core ontology (SAREF) for specification and implementation of complex IoT solutions steered by use cases and requirements aligned with the knowledge represented in and representable by ontology.

When using such a ontology, if additional technologies are to be utilized for data acquisition from the user, like smartphones or embedded solutions, the overall solution takes on characteristics of the IIoT instead of simpler IoT. This additionally raises the importance of non-functional requirements, which are inherent to the OT environments.

For manipulating the existing SAREF4EHAW extension and adding new 'wbah' relations and attributes to it, free and open-source Protégé tool has been used \cite{b56}, which supports OWL2 Web Ontology Language and various formats for importing and exporting ontologies. Further procedural scrutiny is needed to build the new SAREF4EHAW extension candidate, covering all WBAH aspects envisioned by the ontological requirements and supporting all WBAH dimensions with the appropriate level of knowledge structure specified in the new WBAH ontology.

It is important to emphasize that sensors are the basic building elements of any IoT, and their measurement parameters, as well as availability under commercial or laboratory terms, dictate how purposeful the designed IoT ontology will be. As stated in the comprehensive study, "within many wearable electronic products, it is the sensors which provide the key value proposition" so they covered "17 different types of sensor, across 10 major categories, characterising the technology, applications, and industry landscape for this", as follows \cite{b57}: (1) IMUs - inertial measurement units; (2) Optical; (3) Cameras; (4) Electrodes; (5) Force, pressure and stretch; (6) Temperature; (7) Microphones; (8) GPS; (9) Chemical and gas; (10) Others. Therefore, classification of sensors with their particular characteristics and measurement parameters need to be modeled into the targeted ontology.

After the new WBAH ontology is designed and fully specified, verification and validation have to be conducted in order to check the ontology for specification errors and rigor, as well as the appropriateness towards WBAH use cases.

To go another step further, inspired by the concepts of "one health" \cite{b58} and "one digital health" \cite{b59}, as well as the learnings from the current global COVID-19 pandemic \cite{b60}, the notion of "one well-being, aging and health" (OWBAH) can be introduced, covering WBAH aspects of all living beings and their engagement and interactions within communities and ecosystems. Giving the transdisciplinary nature of this concept, the need for tighter collaboration of human medicine, veterinary medicine, environmental health, public health, and the social sciences is evident, in order to prevent and mitigate future health crises caused by novel infectious diseases as well as provide well-being of people, animals, plants, and their shared environment.

The study on human-animal companionship \cite{b61} reviewed published research from 1980 to 2013 undertaken in the field of companion animals and the health of older people, considering "the impact on the physical, psychological, emotional and social health of older people, both in the community as pet owners and as residents of care facilities and other institutions to whom animals are introduced for recreational and therapeutic purposes". The review showed "both the extensive and therapeutic benefits to elderly people provided by pets and companion animals, and the associated positive social and economic influences for local communities and society as a whole" while also recognized some drawbacks, like some older people neglecting their own health, avoid seeking medical care or resist
medical advice because of their companion animal, as well as ignoring advice to find another home for their pet because of allergies.

The global initiative for age-friendly environments was launched in 2007 by the WHO \cite{b62}, inviting cities and communities to sign up with WHO's Global Network of Age-friendly Cities and Communities, which included 1333 cities and communities in 47 countries, covering over 298 million people worldwide in period 2010- 2022. Age-friendly environments, recognized as Smart Healthy Age-Friendly Environments (SHAFE) \cite{b63}, offer an integral approach on eight domains of life: (1) Outdoor spaces and buildings; (2) Transport and mobility; (3) Housing; (4) Social participation; (5) Social inclusion and non-discrimination; (6) Civic engagement and employment; (7) Communication and information; (8) Community and health services.

\section{Conclusion}
Achieving advancements in well-being, aging and health of individuals, communities and ecosystems requires a multidisciplinary effort in order to ensure a more compassionate and healthy age-friendly society. The concept of One Well-Being, Aging and Health, including these aspects for people, animals, plants, and their shared environment, adds even more complexity to the efforts.

If supported by ontologies as advanced means from knowledge engineering, more areas of research and development of WBAH solutions may benefit, e.g. requirements engineering and IoT-based design. Market trends for wearable sensors production show there will be even more "made for wearable" products available, and in order to support their implementation in the future WBAH services, the ontology design should be modular and extensible, and the development process as collaborative and transdisciplinary as possible.

Future work will include conducting the scoping review for the IIoT-based ontology development for WBAH, elaborating on the usage of ontologies for implementation of IoT solutions, aiming at extending the ontology of choice, reusing parts of existing ontologies and adding new constructs needed for cross-domain reasoning. In order to enable more efficient use of resources at the application level, such an ontology should represent an abstract layer of semantic middleware which should reason on acquired sensor data and allow for application business logic to trigger actuators while preserving non-functional properties, such as safety and security.

\section*{Acknowledgment}
This paper is based upon work from COST Action CA16226 - Indoor Living Space Improvement: Smart Habitat for the Elderly (SHELD-ON) and COST Action CA19136 - International Interdisciplinary Network on Smart Healthy Age-friendly Environments (NET4AGE-FRIENDLY), both supported by COST (European Cooperation in Science and Technology). Part of this work has been done within the "Social Innovation for integrated health CARE of ageing population in ADRION Regions (SI4CARE)" project, supported by the European Regional Development Fund (ERDF), and the "Support for the development of the Croatian e-Health Strategic Development Plan 2020-2025 and Action Plan 2020-2021" (Request for Service ID: SRSS/SC2019/163), supported by The European Union’s SRSS/2018/01/FWC/002 in EU Member States.

\end{document}